\documentclass{article}
\usepackage[utf8]{inputenc}
\usepackage{arxiv,amsmath,amssymb,bm,bbm,mathtools,braket,hyperref,url,xcolor}

\usepackage[linesnumbered,ruled]{algorithm2e}

\SetCommentSty{mycommfont}

\newcommand{\iu}{{i\mkern1mu}}

\newcommand{\rhoin}{\ket{\text{in}}\!\bra{\text{in}}}

\usepackage{tocloft}
\title{Efficient calculation of gradients in classical simulations of variational quantum algorithms}
\author{
    Tyson Jones 
    \\
    IBM Research UK \\
    The Hartree Centre, Daresbury Laboratory \\ Warrington, WA4 4AD, UK \\
    \texttt{tyson.jones@ibm.com} \\
    \And
    Julien Gacon \\
    IBM Quantum, IBM Research -- Zurich \\ 8803 Rueschlikon, Switzerland \\
    \texttt{jul@zurich.ibm.com}
}

\begin{document}
\maketitle

\begin{abstract}
    Calculating the energy gradient in parameter space has become an almost ubiquitous subroutine of variational near-term quantum algorithms~\cite{preskill_nisq,yuan_theoryofvariational,harrow2019low}. ``Faithful'' classical emulation of this subroutine mimics its quantum evaluation~\cite{yaojl_vari}, and scales as $\mathcal{O}(P^2)$ gate operations for $P$ variational parameters. This is often the bottleneck for the moderately-sized simulations, and has attracted HPC strategies like ``batch-circuit" evaluation~\cite{google_tensorflow_quantum,intel_qhipster_vari}.
    We here present a novel derivation of an emulation strategy to precisely calculate the gradient in $\mathcal{O}(P)$ time and using $\mathcal{O}(1)$ state-vectors, compatible with ``full-state" state-vector simulators.
    The prescribed algorithm resembles the optimised technique for automatic differentiation of reversible cost functions~\cite{margossian2019review}, often used in classical machine learning~\cite{baydin2017automatic}, and first employed in quantum simulators like \texttt{Yao.jl}~\cite{yaojl_vari}.
    In contrast, our scheme derives directly from a recurrent form of quantum operators,
    and may be more familiar to a quantum computing community.
    Our strategy is very simple, uses only ``apply gate", ``clone state" and ``inner product" primitives and is hence straightforward to implement and integrate with existing simulators. It is compatible with gate parallelisation schemes, and hardware accelerated and distributed simulators.
    We describe the scheme in an instructive way, including details of how common gate derivatives can be performed, to clearly guide implementation in existing quantum simulators.
    We furthermore demonstrate the scheme by implementing it in Qiskit~\cite{cross2018ibm,Qiskit_big_ref}, and perform some comparative benchmarking with faithful simulation.
    Finally, we remark upon the difficulty of extending the scheme to density-matrix simulation of noisy channels.
\end{abstract}


\section{Introduction}

Variational quantum algorithms show promise as an early application of near-term and noisy intermediate-scale quantum (NISQ) computers~\cite{preskill_nisq,peruzzo2014variational}.
They involve iteratively producing a succession of quantum states from a parameterized ``ansatz" circuit, in order to find the optimum of some measurable cost function.
Many employ a gradient based optimiser~\cite{yuan_theoryofvariational,harrow2019low}, whereby the gradient of the cost function is evaluated, with respect to the ansatz parameters, and is used in updating them.
For example, gradient descent to find the ground-state energy under a Hamiltonian prescribes a change in parameters $\vec\theta$ of
\begin{align}
 \Delta \vec\theta \propto - \nabla \braket{E(\vec\theta)},
\end{align}
where $\braket{E(\vec\theta)}$ is the expected energy of the ansatz state informed by $\vec \theta$, 
and
where each gradient entry $\partial \braket{E}/\partial \theta_i$ is evaluated  independently. There are several techniques to perform this evaluation~\cite{PhysRevA.99.032331,PhysRevLett.118.150503,bergholm2018pennylane,mcardle2019variational}, with similar quantum resource costs; For $P = \dim \vec\theta\;$ parameters, estimating the energy gradient will involve $\mathcal{O}(P^2)$ gates total, excluding measurements.
A simple illustration is via finite-difference approximation,
\begin{align}
    \delta \theta \, \frac{\partial \braket{E}}{\partial \theta_i} \approx 
    \braket{E(\theta_i + \delta \theta)} - \braket{E(\theta_i)},
\end{align}
where evaluating each $\braket{E(\theta_i + \delta \theta)}$ requires a fixed number of evaluations of the full $\Omega(P)$-gate ansatz.

Classical simulation of variational quantum algorithms like these is a crucial step in their development. Owing to the exponentially growing cost of simulating even a perfect quantum computer, considerable effort has been invested in building high-performance and hardware-accelerated simulators~\cite{yaojl_vari,cross2018ibm,Qiskit_big_ref,jones2019quest,jones2020questlink,steiger2018projectq}.
These simulators aim to speedup simulation of general quantum circuits at the gate level.
Ergo, they can classically evaluate the energy gradient by a direct emulation of the quantum evaluation, in $\mathcal{O}(P^2)$ gates.
This turns out to be a sub-optimal parallelisation granularity for simulating variational algorithms. %
Very recently, so-called ``batch" strategies have emerged for parallel evaluation of entire circuits, which can in combination speed up simulation of variational routines~\cite{google_tensorflow_quantum,intel_qhipster_vari}.
Though they admit the same $\mathcal{O}(P^2)$ scaling, they can use parallel hardware to, for example, simultaneously evaluate ${\partial} \braket{E}/{\partial} \theta_i$ for several values of $i$.

Since ansatz circuits are typically unitary, and since unitaries are reversible, an asymptotically faster strategy is possible. The so-called ``reverse mode" of automatic differentiation~\cite{margossian2019review,bartholomew2000automatic}, a canonical technique for classically evaluating gradients of cost functions in the machine learning literature~\cite{baydin2017automatic}, scales in runtime as $\mathcal{O}(P)$. This technique usually involves caching intermediate states of the evaluation, at a multiplicative $P$ cost in memory~\cite{margossian2019review}, though this can be reduced to a constant overhead for reversible cost functions~\cite{maclaurin2015gradient}. Indeed, this has been employed for speeding up evaluation of quantum gradients in \texttt{Yao.jl}~\cite{yaojl_vari}, a recent state-of-the-art quantum simulator with leading performance in simulating variational algorithms.

In this technical note, we present a similar technique with the same runtime and memory costs, derived directly from a recurrency in the analytic form of the gradient. 
It prescribes a simple re-ordering of how the analytic forms of the energy derivatives are numerically evaluated, to avoid repeated simulation of any one ansatz gate.
We outline how it can precisely compute the entire gradient $\nabla \braket{E}$ in $\mathcal{O}(P)$ gate primitives, and $\mathcal{O}(1)$ memory, without invoking caching or finite-difference approximations. We describe in detail how the technique can be integrated into existing quantum simulators, and even discuss how gate derivatives can be enacted with existing simulator facilities in Appendix~\ref{app:gate_derivs}. We also present extensions to the algorithm to support multi-parameter gates, non-unique ansatz parameters, non-unitary ansatz circuits and non-Hermitian cost operators, in Appendix~\ref{app:state_vec_extensions}.
We stress that our algorithm is a \textit{strong-simulation} strategy, rather than one for emulation, and hence is most useful to quickly obtain the behaviour of a gradient-based algorithm when run on a perfect quantum machine. 

\section{Gradient evaluation}

\subsection{Scope}


Below, we detail our simulation strategy for efficient classical evaluation of gradients of any expected value, though we use energy under a Hamiltonian as an example. We make no assumption about the form of the Hamiltonian --- it can be time dependent, and may change freely between evaluations of the gradient. Even the condition of Hermiticity can be relaxed, as shown in Appendix~\ref{app:non_herm_op}.
For simplicity, the outline of our algorithm below assumes each gate has a single unique parameter, though our scheme is easily extended to repeated parameters and multi-parameter gates, as presented in Appendices~\ref{app:multi_param} and \ref{app:repeat_param}. Our presentation assumes the ansatz circuit is unitary, but this may also be relaxed, as discussed in Appendix~\ref{app:non_unitary_ansatz}.
In its current form, our strategy applies only for noise-free state-vector simulation, though we discuss the seemingly less permissive task of density matrix simulation in Appendix~\ref{app:dens_matr}. 

We make few assumptions about the capability of the simulator. We require it can apply the operator of interest, e.g. Hamiltonian, to a state-vector and hence produce an intermediate unnormalised state. Note even a non-Hermitian operator is compatible, to admit an imaginary gradient. We assume a non-normalised state can have further gates operated upon it, and can have its inner product with another state calculated. We assume applying inverse unitaries is supported and efficient, as is applying the derivative of a gate, and we outline how to compute such derivatives in Appendix~\ref{app:gate_derivs}. Note a gate derivative is in general non-unitary, and need not be calculated analytically; it can be evaluated numerically with e.g. finite difference methods. We hence furthermore assume the simulator can multiply non-unitary but tractable matrices upon a state-vector.
These facilities are simple and present in practically all modern quantum computing simulation frameworks.
By using \textit{only} these assumed facilities, our algorithm is compatible with other parallelisation and optimisation schemes used by high-performance simulators, like hardware acceleration and distribution.

\subsection{Derivation}

We here derive an analytic recurrence relation for the gradient. Understanding the derivation is an important step in understanding the subsequent algorithm.
Let $\braket{E(\vec\theta)}$ be the energy (under Hamiltonian $\hat{H}$) of a pure state $\ket{\psi(\vec\theta)}$, produced by a parameterized ansatz circuit $\hat{U}(\vec\theta)$ acting on fixed input state $\ket{\text{in}}$. That is
\begin{align}
    \ket{\psi(\vec\theta)} = \hat{U}(\vec\theta) \ket{\text{in}}.
\end{align}

The $i$-th element of the energy gradient is
\begin{align}
\frac{\partial\braket{E}}{\partial \theta_i} 
&=
\frac{\partial}{\partial \theta_i} 
\braket{\psi(\vec\theta)|\hat{H}|\psi(\vec\theta)} 
=
\frac{\partial}{\partial \theta_i} 
\braket{\text{in}| \hat U^\dagger(\vec\theta)\hat{H}\hat U(\vec\theta) |\text{in}}
\\
&=
\bra{\text{in}} \frac{\partial \hat U^\dagger(\vec\theta)}{\partial \theta_i} \hat H \hat U(\vec\theta) \ket{\text{in}}
+
\bra{\text{in}} \hat U^\dagger(\vec\theta)  \hat H  \frac{\partial \hat U(\vec\theta)}{\partial \theta_i} \ket{\text{in}}
\\
&= 2 \; \Re \;
			\bra{\text{in}} \hat U^\dagger(\vec\theta) \hat{H}  \frac{\partial \hat U(\vec\theta)}{\partial \theta_i} \ket{\text{in}},
\label{eq:deriv_h_hermitian_step}
\end{align}
invoking $\hat{H} = \hat{H}^\dagger$.
Assume the ansatz $\hat{U}$ is composed of $P$ gates, $\hat{U}_i$, each with a unique parameter $\theta_i$. That is, $\hat{U}(\vec\theta) = \hat{U}_P(\theta_P) \dots \hat{U}_1(\theta_1)$.
Then
\begin{align}
\frac{\partial\braket{E}}{\partial \theta_i}
    &= 2 \; \Re \;
			\bra{\text{in}} 
			\hat U^\dagger_1(\theta_1) 
			\dots 
			\hat U^\dagger_P(\theta_P)
			\;
			\hat{H} 
			\;
			\hat U_P(\theta_P) 
			\dots 
			\frac{\mathrm{d}\hat{U}_i}{\mathrm{d} \theta_i}
			\dots 
			\hat U_1(\theta_1) 
			\ket{\text{in}}.
\end{align}
Notate
\begin{align}
    U_{i:j} &= \prod\limits_{k=j}^i \hat{U}_k(\theta_k),
    &
    &\text{and}
    &
    \text{prod}\big[ \ket{a}, \; \ket b \big] &= \braket{a|b}.
\end{align}
We add no hat to symbol $U_{i:j}$ merely to emphasise it as a sequence of gate primitives, rather than a single gate.
The $i$-th element of the gradient can then be expressed as
\begin{align}
    \frac{\partial \braket{E}}{\partial \theta_i}
    &=
    2 \; \Re \; 
    \text{prod}\big[  U_{1:P} \, \ket{\text{in}}, \;
    \; \hat{H} \; 
    U_{i+1:P} 
    \; \frac{\mathrm{d}\hat{U}_i}{\mathrm{d} \theta_i} \;
    U_{1:i-1} \;
    \ket{\text{in}}
    \big],  \label{eq:reference_gradient}
    \\
    &=
    2 \; \Re \; 
    \text{prod}\big[ {U_{i+1:P}}^\dagger
    \;   \hat{H} \;  U_{1:P} \, \ket{\text{in}}, \;
    \; 
    \frac{\mathrm{d}\hat{U}_i}{\mathrm{d} \theta_i} \;
    {U_{i:P}}^\dagger \; U_{1:P} \;
    \ket{\text{in}}
    \big],
    \\
    &=
    2 \; \Re \; 
    \text{prod}\big[ {U_{i+1:P}}^\dagger
    \;   \hat{H} \,  \ket{\psi}, \;
    \; 
    \frac{\mathrm{d}\hat{U}_i}{\mathrm{d} \theta_i} \;
    {U_{i:P}}^\dagger \ket{\psi}
    \big].
\end{align}
By denoting
\begin{gather}
    \ket{\phi}_i = {U_{i:P}}^\dagger \ket{\psi}
    \; \; \; \implies 
    \ket{\phi}_{i} = \hat{U}_{i}^\dagger \ket{\phi}_{i+1}, \;\;\; 
    \\
    \ket{\lambda}_{i} = {U_{i+1:P}}^\dagger
    \;   \hat{H} \,  \ket{\psi} \; \; \; \implies \ket{\lambda}_i = \hat{U}_{i+1}^\dagger \ket{\lambda}_{i+1},
\end{gather}
we make explicit the recurrence leveraged by our scheme;
\begin{align}
    \frac{\partial \braket{E}}{\partial \theta_i}
    &=
    2 \; \Re \; \text{prod} \big[ 
    \ket{\lambda}_{i}, \;\;
     \frac{\mathrm{d}\hat{U}_i}{\mathrm{d} \theta_i} \;
     \ket{\phi}_i
    \big]
    \label{eq:recurrent_grad}
    \\
    &=
    2 \; \Re \; \text{prod} \big[ 
    \hat{U}_{i+1}^\dagger \ket{\lambda}_{i+1}, \;\;
     \frac{\mathrm{d}\hat{U}_i}{\mathrm{d} \theta_i} \;
     \hat{U}_i^\dagger
     \ket{\phi}_{i+1}
    \big].
\end{align}

\subsection{Algorithm}

The algorithm is a simple reordering of the operations involved in numerically evaluating the analytic form of the gradient, by the recurrence relationship derived above. That is, we evaluate Equation~\ref{eq:recurrent_grad}, from $i=P$ to $i=1$, where $\ket{\lambda}_i$ and $\ket{\phi}_i$ are iteratively procured from their previous assignment. This avoids applying each gate in the ansatz more than a fixed number of times. We formally present our strategy in Algorithm~\ref{alg:state_vec_grad}. 

{\centering 
\begin{minipage}{.7\linewidth}
\begin{algorithm}[H]
\DontPrintSemicolon
\SetKwInOut{Input}{Input}
\SetKwInOut{Output}{Output}

\Input{State-vectors $\ket{\lambda}$, $\ket{\phi}$, $\ket{\mu}$, an immutable input state $\ket{\text{in}}$, some representation of a circuit $U_{1:P}$ with a single unique parameter in each gate, and a Hamiltonian $\hat{H}$ in any applicable representation}
\Output{Each element of $\nabla \braket{E}$}

$ \ket{\lambda} \coloneqq  \ket{\text{in}}$ 
\tcp*{clone state in $\mathcal{O}(G)$}

$ \ket{\lambda} \leftarrow \hat{U}_{1:P} \ket{\lambda}$
\tcp*{apply $P$ gates in $\mathcal{O}(P\,  G)$}

$\ket{\phi} \coloneqq \ket{\lambda} $
\tcp*{clone state in $\mathcal{O}(G)$}

$ \ket{\lambda} \leftarrow \hat{H} \ket{\lambda}$
\tcp*{apply $\hat{H}$ in $\mathcal{O}(h \, N \, G)$}

\For{$i \in \{P, \dots, 1\}$} 
  {
    $\ket{\phi} \leftarrow \hat{U}_i^\dagger \ket{\phi}$ 
    \tcp*{apply gate in $\mathcal{O}(G)$}
    	            
    $\ket{\mu} \coloneqq \ket{\phi}$
    \tcp*{clone state in $\mathcal{O}(G)$}
    
    $\ket{\mu} \leftarrow (\mathrm{d} \hat{U}_i / \mathrm{d}\theta_i ) \ket{\mu}$ 
    \tcp*{apply non-unitary in $\mathcal{O}(G)$}
    
    $\nabla\braket{E}_i = 2 \; \Re \; \braket{\lambda | \mu}$
    \tcp*{compute inner product in $\mathcal{O}(G)$}
    
    \If{$i>1$}
    {
    	$ \ket{\lambda} \leftarrow U_i^\dagger \ket{\lambda}$
    	\tcp*{apply gate in $\mathcal{O}(G)$}
    }
  } 
\caption{Calculating the noise-free gradient with state-vectors, using ``reverse mode". Let $G$ be the complexity of effecting a fixed-size gate upon an $N$-qubit state-vector. Typically $G$ scales with the number of amplitudes in the state-vector as $G = \mathcal{O}(2^N)$.}
\label{alg:state_vec_grad}
\end{algorithm}
  \end{minipage}
  \par }
  
The total number of gates simulated in Algorithm~\ref{alg:state_vec_grad} for an $N$-qubit $P$-parameter ansatz is $\mathcal{O}(P  + h N)$, where $h$ is the number of terms in the Hamiltonian.
An additional $P$ inner products are performed, though each is typically as costly as a single gate.
In general, the once-off and unavoidable cost of applying the Hamiltonian will involve strictly fewer than $h N$ gate operations, depending on its representation. Despite Hamiltonians in the Pauli basis permitting $h = \mathcal{O}(4^N)$ terms, typical tractable Hamiltonians of interest grow polynomially, like $\mathcal{O}(N^4)$~\cite{peruzzo2014variational}. Note also that the cost of evaluating $\hat{H}\ket{\lambda}$ in Line~4 of Algorithm~\ref{alg:state_vec_grad} is likely already paid during simulation, in order to compute the expected energy of the current parameter assignment. Hence, computing the full energy gradient via our algorithm costs $\mathcal{O}(P)$ gate primitives, and $\mathcal{O}(1)$ additional memory.
We here-from loosely refer to our algorithm as "reverse mode", to distinguish it from faithful techniques of gradient estimation.

\section{Benchmarking}

We benchmark a new Qiskit implementation of Algorithm~\ref{alg:state_vec_grad}, and compare it to a reference gradient computation using the representation from Equation~\ref{eq:reference_gradient}.
Since each of the $P$ gradient entries in $\nabla \braket{E}$ requires applying $P$ gates, our reference calculation scales as $\mathcal{O}(P^2)$.
The reference algorithm is made explicit in Algorithm~\ref{alg:state_vec_grad_standard}.

{\centering 
\begin{minipage}{.7\linewidth}
\begin{algorithm}[H]
\DontPrintSemicolon
\SetKwInOut{Input}{Input}
\SetKwInOut{Output}{Output}

\Input{State-vectors $\ket{\lambda}$, $\ket{\mu}$, an immutable input state $\ket{\text{in}}$, some representation of a circuit $U_{1:P}$ with a single unique parameter in each gate, and a Hamiltonian $\hat{H}$ in any applicable representation}
\Output{Each element of $\nabla \braket{E}$}

$ \ket{\lambda} \coloneqq  \ket{\text{in}}$ 
\tcp*{clone state in $\mathcal{O}(G)$}

$ \ket{\lambda} \leftarrow \hat{U}_{1:P} \ket{\lambda}$
\tcp*{apply $P$ gates in $\mathcal{O}(P\,  G)$}

$ \ket{\lambda} \leftarrow \hat{H} \ket{\lambda}$
\tcp*{apply $\hat{H}$ in $\mathcal{O}(h \, N \, G)$}

\For{$i \in \{1, \dots, P\}$} 
  {
    $ \ket{\mu} \coloneqq  \ket{\text{in}}$ 
    \tcp*{clone state in $\mathcal{O}(G)$}
    
    $\ket{\mu} \leftarrow \hat{U}_{1:i-1} \ket{\mu}$ 
    \tcp*{apply $i-1$ gates in $\mathcal{O}((i-1) \, G)$}
    
    $\ket{\mu} \leftarrow (\mathrm{d} \hat{U}_i / \mathrm{d}\theta_i ) \ket{\mu}$ 
    \tcp*{apply non-unitary in $\mathcal{O}(G)$}
    
    $\ket{\mu} \leftarrow \hat{U}_{i+1:P} \ket{\mu}$ 
    \tcp*{apply $P-i-1$ gates in $\mathcal{O}((P-i-1) \, G)$}
    	            
    $\nabla\braket{E}_i = 2 \; \Re \; \braket{\lambda | \mu}$
    \tcp*{compute inner product in $\mathcal{O}(G)$}
    
  } 
\caption{Calculating the noise-free gradient with state-vectors using a standard gradient scheme.}
\label{alg:state_vec_grad_standard}
\end{algorithm}
  \end{minipage}
  \par }

\begin{figure}
    \centering
    \begin{minipage}{0.19\textwidth}
        \centering
        \includegraphics[height=2.5cm]{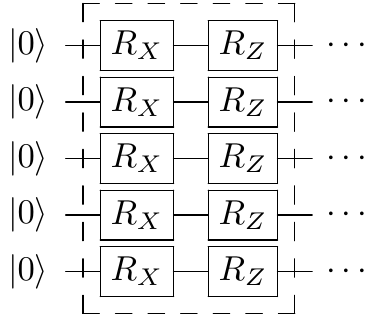}
        \\
        Circuit A
    \end{minipage}
    \begin{minipage}{0.35\textwidth}
        \centering
        \includegraphics[height=2.5cm]{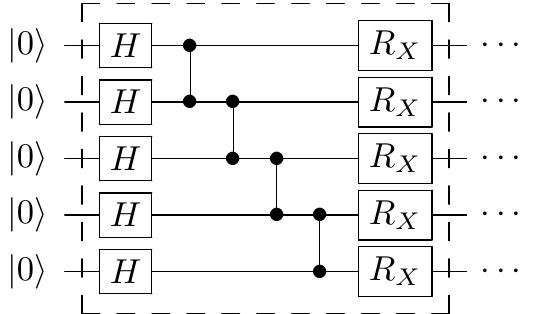}
        \\
        Circuit B
    \end{minipage}
    \begin{minipage}{0.45\textwidth}
        \centering
        \includegraphics[height=2.5cm]{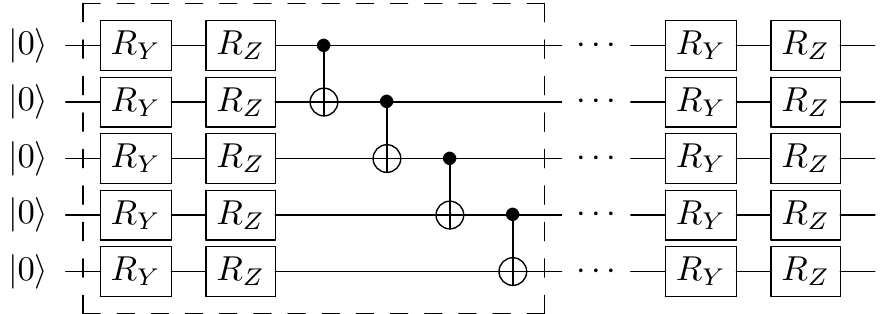}
        \\
        Circuit C
    \end{minipage}
    
    \begin{minipage}{\textwidth}
        \mbox{~} \\
    \end{minipage}
    
    \begin{minipage}{0.8\textwidth}
        \centering
        \includegraphics[height=2.1cm]{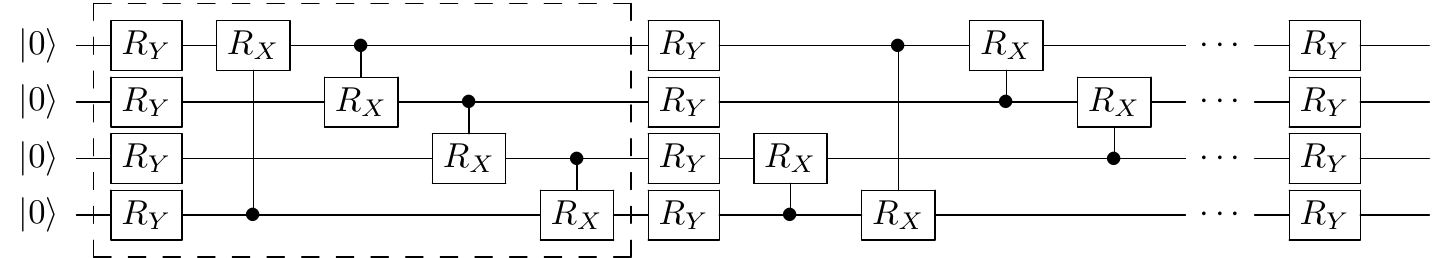}
        \\
        Circuit D
    \end{minipage}
    \caption{Ansatz circuits used in comparitive benchmarking of Algorithms~\ref{alg:state_vec_grad}~and~\ref{alg:state_vec_grad_standard}.
    The dashed region is repeated to increase the number of parameters during testing; everything after the dots is a fixed circuit suffix.
    Circuits A, B and D are chosen from Reference~\cite{sim2019ansatze} for their low and high expressibilities respectively. Note that Circuit D has a slightly different entanglement structure as originally proposed, namely every entanglement layer swaps the role of control and target qubit and periodically shifts all controlled-$R_X$ gates. 
    To clarify this, the Figure of Circuit D also shows the \textit{second} layer with altered structure.
    Circuit C is the hardware efficient SU(2) 2-local circuit provided by Qiskit~\cite{Qiskit_big_ref}.
    }
    \label{fig:ansatz_circuits}
\end{figure}
\begin{figure}
    \centering
    \begin{minipage}{0.45\textwidth}
        \centering
        \includegraphics[width=\textwidth]{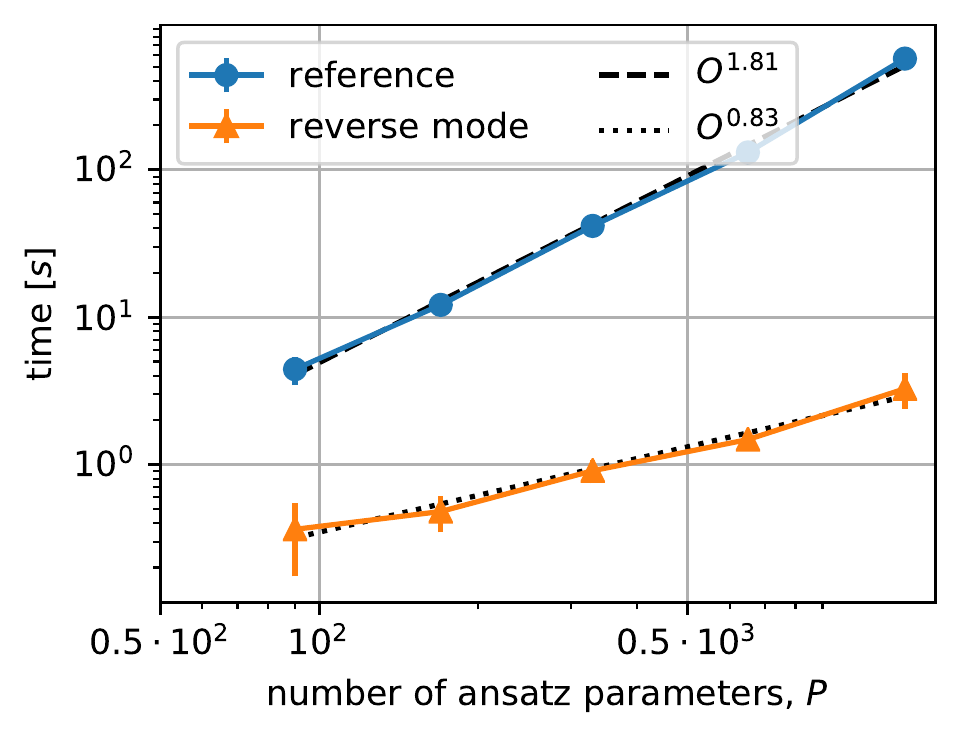}
        \\
        Circuit A
    \end{minipage}
    \begin{minipage}{0.45\textwidth}
        \centering
        \includegraphics[width=\textwidth]{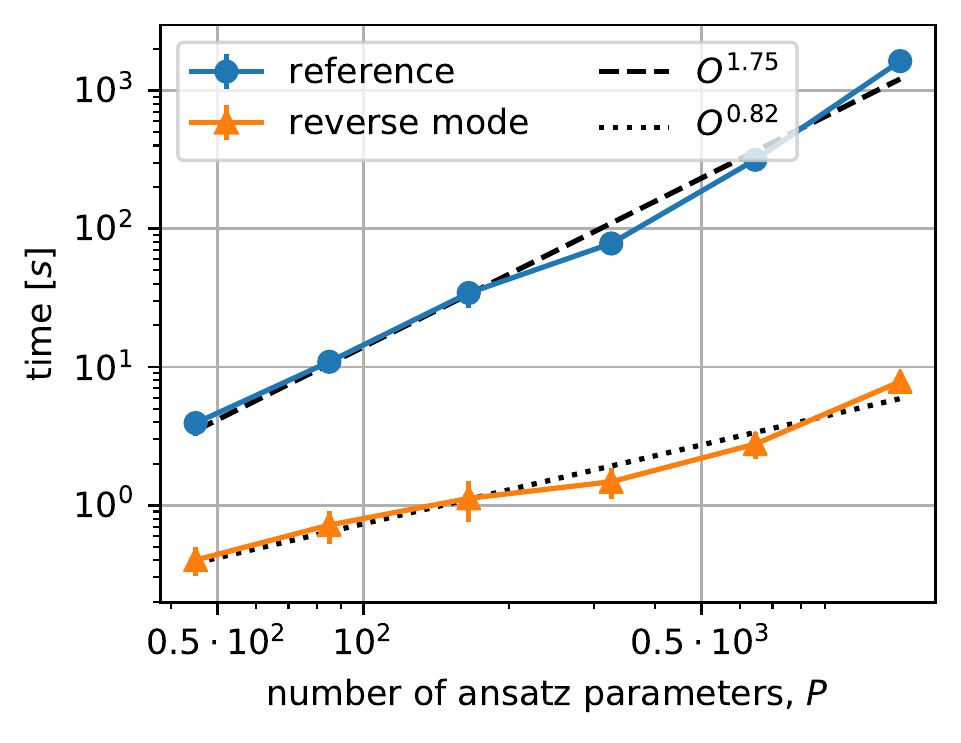}
        \\
        Circuit B
    \end{minipage}
        \begin{minipage}{0.45\textwidth}
        \centering
        \includegraphics[width=\textwidth]{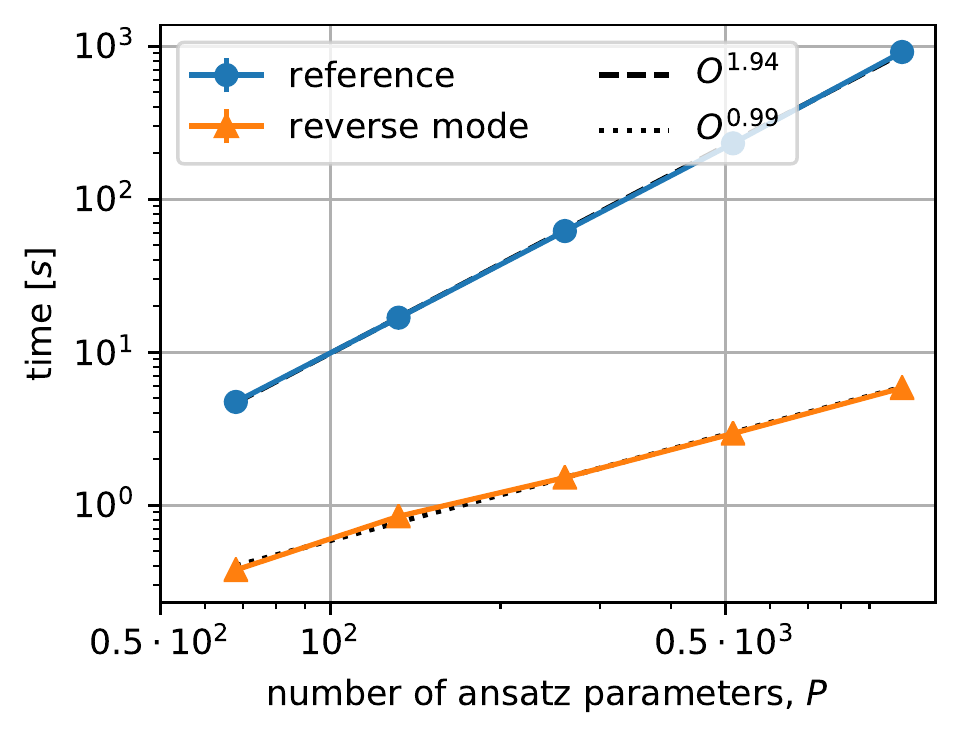}
        \\
        Circuit C
    \end{minipage}
    \begin{minipage}{0.45\textwidth}
        \centering
        \includegraphics[width=\textwidth]{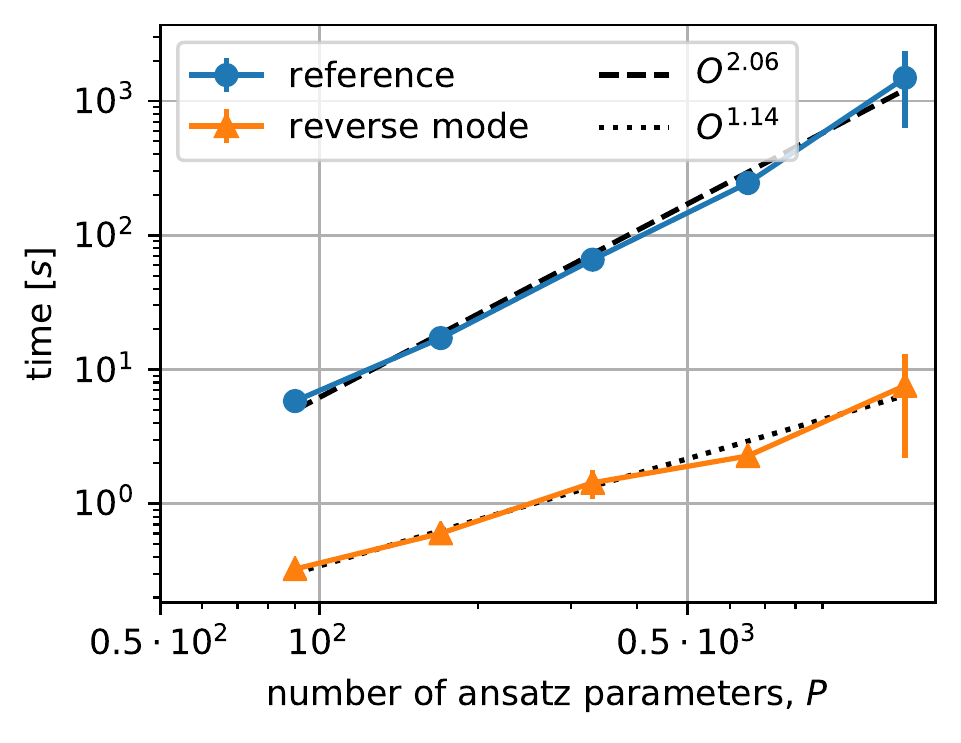}
        \\
        Circuit D
    \end{minipage}
    \caption{Runtime comparison of Algorithms~\ref{alg:state_vec_grad}~and~\ref{alg:state_vec_grad_standard}, referred to as ``reverse mode" and ``reference" respectively,
    calculating the gradient of the ansatz circuits presented in Figure~\ref{fig:ansatz_circuits}.
    Each point shows the average and standard deviation of $24$ runs. 
    The dashed lines are linear fits to the logarithm of the runtime 
    versus the logarithm of the number of parameters.
    }
    \label{fig:runtime_benchmarks}
\end{figure}

We benchmark the two schemes computing the gradients of four structurally distinct classes of ansatz circuits, shown in Figure~\ref{fig:ansatz_circuits}. This includes circuits nominated for their expressibility and entangling capability~\cite{sim2019ansatze}, as well as the hardware efficient SU(2) 2-local circuit provided by Qiskit~\cite{Qiskit_big_ref}. This latter circuit is a heuristic pattern, and a good representation of a typical ansatz circuit used in the literature~\cite{mcardle2019variational,kandala2017hardware,PhysRevA.99.062304}.
For each class of circuits, we vary the number of parameters and resulting circuit depth up to $P=1290$, and measure the runtime of both algorithms to compute the full gradient under a simple Hamiltonian $\hat{H} = \text{Hadamard}^{\otimes N}$.
We fix the number of qubits (at $N = 4, 5$), so as to fix the cost of the ``apply gate", ``clone state" and ``inner product" operations.
The simulation results are presented in Figure~\ref{fig:runtime_benchmarks}, and show excellent agreement with the expected $\mathcal{O}(P)$ speedup of our reverse mode over the reference gradient calculation. 

\subsection{Code availability}
  
The benchmark code is available at \url{github.com/Cryoris/gradient-reverse-mode}.

\section{Author Contributions}

TJ devised the algorithm and wrote the manuscript. JG implemented the algorithm in Qiskit and performed and presented the benchmarking.

\section{Acknowledgements}

TJ thanks IBM Research UK for the opportunity to undertake an internship in quantum computing at the Daresbury Laboratory. 
TJ additionally thanks Xiu-Zhe (Roger) Luo and Balint Koczor for helpful and insightful discussions.

IBM, the IBM logo, and ibm.com are trademarks of International Business Machines Corp., registered in many jurisdictions worldwide. Other product and service names might be trademarks of IBM or other companies. The current list of IBM trademarks is available at \url{https://www.ibm.com/legal/copytrade}.

\bibliographystyle{unsrt}
\bibliography{biblio.bib}

\begin{thebibliography}{10}

\bibitem{preskill_nisq}
John Preskill.
\newblock Quantum {C}omputing in the {NISQ} era and beyond.
\newblock {\em {Quantum}}, 2:79, August 2018.

\bibitem{yuan_theoryofvariational}
Xiao Yuan, Suguru Endo, Qi~Zhao, Ying Li, and Simon~C. Benjamin.
\newblock Theory of variational quantum simulation.
\newblock {\em {Quantum}}, 3:191, October 2019.

\bibitem{harrow2019low}
Aram Harrow and John Napp.
\newblock Low-depth gradient measurements can improve convergence in
  variational hybrid quantum-classical algorithms.
\newblock {\em arXiv preprint arXiv:1901.05374}, 2019.

\bibitem{yaojl_vari}
Xiu-Zhe Luo, Jin-Guo Liu, Pan Zhang, and Lei Wang.
\newblock Yao.jl: Extensible, efficient framework for quantum algorithm design,
  2019.

\bibitem{google_tensorflow_quantum}
Michael Broughton, Guillaume Verdon, Trevor McCourt, Antonio~J. Martinez,
  Jae~Hyeon Yoo, Sergei~V. Isakov, Philip Massey, Murphy~Yuezhen Niu, Ramin
  Halavati, Evan Peters, Martin Leib, Andrea Skolik, Michael Streif, David~Von
  Dollen, Jarrod~R. McClean, Sergio Boixo, Dave Bacon, Alan~K. Ho, Hartmut
  Neven, and Masoud Mohseni.
\newblock Tensorflow quantum: A software framework for quantum machine
  learning, 2020.

\bibitem{intel_qhipster_vari}
Gian~Giacomo Guerreschi, Justin Hogaboam, Fabio Baruffa, and Nicolas Sawaya.
\newblock Intel quantum simulator: A cloud-ready high-performance simulator of
  quantum circuits, 2020.

\bibitem{margossian2019review}
Charles~C Margossian.
\newblock A review of automatic differentiation and its efficient
  implementation.
\newblock {\em Wiley Interdisciplinary Reviews: Data Mining and Knowledge
  Discovery}, 9(4):e1305, 2019.

\bibitem{baydin2017automatic}
At{\i}l{\i}m~G{\"u}nes Baydin, Barak~A Pearlmutter, Alexey~Andreyevich Radul,
  and Jeffrey~Mark Siskind.
\newblock Automatic differentiation in machine learning: a survey.
\newblock {\em The Journal of Machine Learning Research}, 18(1):5595--5637,
  2017.

\bibitem{cross2018ibm}
Andrew Cross.
\newblock The {IBM Q} experience and {QISKit} open-source quantum computing
  software.
\newblock {\em APS}, 2018:L58--003, 2018.

\bibitem{Qiskit_big_ref}
{Qiskit} contributors.
\newblock Qiskit: An open-source framework for quantum computing, 2019.

\bibitem{peruzzo2014variational}
Alberto Peruzzo, Jarrod McClean, Peter Shadbolt, Man-Hong Yung, Xiao-Qi Zhou,
  Peter~J Love, Al{\'a}n Aspuru-Guzik, and Jeremy~L O’brien.
\newblock A variational eigenvalue solver on a photonic quantum processor.
\newblock {\em Nature communications}, 5:4213, 2014.

\bibitem{PhysRevA.99.032331}
Maria Schuld, Ville Bergholm, Christian Gogolin, Josh Izaac, and Nathan
  Killoran.
\newblock Evaluating analytic gradients on quantum hardware.
\newblock {\em Phys. Rev. A}, 99:032331, Mar 2019.

\bibitem{PhysRevLett.118.150503}
Jun Li, Xiaodong Yang, Xinhua Peng, and Chang-Pu Sun.
\newblock Hybrid quantum-classical approach to quantum optimal control.
\newblock {\em Phys. Rev. Lett.}, 118:150503, Apr 2017.

\bibitem{bergholm2018pennylane}
Ville Bergholm, Josh Izaac, Maria Schuld, Christian Gogolin, Carsten Blank,
  Keri McKiernan, and Nathan Killoran.
\newblock Pennylane: Automatic differentiation of hybrid quantum-classical
  computations.
\newblock {\em arXiv preprint arXiv:1811.04968}, 2018.

\bibitem{mcardle2019variational}
Sam McArdle, Tyson Jones, Suguru Endo, Ying Li, Simon~C Benjamin, and Xiao
  Yuan.
\newblock Variational ansatz-based quantum simulation of imaginary time
  evolution.
\newblock {\em npj Quantum Information}, 5(1):1--6, 2019.

\bibitem{jones2019quest}
Tyson Jones, Anna Brown, Ian Bush, and Simon~C Benjamin.
\newblock Quest and high performance simulation of quantum computers.
\newblock {\em Scientific reports}, 9(1):1--11, 2019.

\bibitem{jones2020questlink}
Tyson Jones and Simon~C Benjamin.
\newblock Questlink--mathematica embiggened by a hardware-optimised quantum
  emulator.
\newblock {\em Quantum Science and Technology}, 2020.

\bibitem{steiger2018projectq}
Damian~S Steiger, Thomas H{\"a}ner, and Matthias Troyer.
\newblock Projectq: an open source software framework for quantum computing.
\newblock {\em Quantum}, 2:49, 2018.

\bibitem{bartholomew2000automatic}
Michael Bartholomew-Biggs, Steven Brown, Bruce Christianson, and Laurence
  Dixon.
\newblock Automatic differentiation of algorithms.
\newblock {\em Journal of Computational and Applied Mathematics},
  124(1-2):171--190, 2000.

\bibitem{maclaurin2015gradient}
Dougal Maclaurin, David Duvenaud, and Ryan Adams.
\newblock Gradient-based hyperparameter optimization through reversible
  learning.
\newblock In {\em International Conference on Machine Learning}, pages
  2113--2122, 2015.

\bibitem{sim2019ansatze}
Sukin Sim, Peter~D Johnson, and Al{\'a}n Aspuru-Guzik.
\newblock Expressibility and entangling capability of parameterized quantum
  circuits for hybrid quantum-classical algorithms.
\newblock {\em Advanced Quantum Technologies}, 2(12):1900070, 2019.

\bibitem{kandala2017hardware}
Abhinav Kandala, Antonio Mezzacapo, Kristan Temme, Maika Takita, Markus Brink,
  Jerry~M Chow, and Jay~M Gambetta.
\newblock Hardware-efficient variational quantum eigensolver for small
  molecules and quantum magnets.
\newblock {\em Nature}, 549(7671):242--246, 2017.

\bibitem{PhysRevA.99.062304}
Tyson Jones, Suguru Endo, Sam McArdle, Xiao Yuan, and Simon~C. Benjamin.
\newblock Variational quantum algorithms for discovering hamiltonian spectra.
\newblock {\em Phys. Rev. A}, 99:062304, Jun 2019.

\bibitem{cai2020resource}
Zhenyu Cai.
\newblock Resource estimation for quantum variational simulations of the
  hubbard model.
\newblock {\em Physical Review Applied}, 14(1):014059, 2020.

\bibitem{non_hermitian_expec_valus}
Arun~Kumar Pati, Uttam Singh, and Urbasi Sinha.
\newblock Measuring non-hermitian operators via weak values.
\newblock {\em Physical Review A}, 92(5):052120, 2015.

\end{thebibliography}
\appendix 

\section{Derivative of a gate}
\label{app:gate_derivs}

Here we present strategies for evaluating the derivative of a parameterised unitary gate, as required in Step~\textbf{8} of Algorithm~\ref{alg:state_vec_grad}.
Rotation gates of the form $\hat{Rx}(\theta) = \exp(\alpha \, \iu \, \theta \hat{X})$, and similarly for $\hat{{Ry}}$ and $\hat{{Rz}}$, admit a simple derivative of the form
\begin{align}
    \frac{\mathrm{d} \hat{{Rx}}(\theta)}{\mathrm{d} \theta} 
    &= \alpha \, i \, \hat{X} \, \hat{{Rx}}(\theta),
\end{align}
and can hence be effected by merely operating both $\hat{X}$ and $\hat{Rx}(\theta)$. The state-vector need not be scaled by coefficient $\alpha \, i$, which can instead be cheaply multiplied with the scalar evaluated in Step \textbf{9} of Algorithm~\ref{alg:state_vec_grad}.
This strategy also holds for rotations around general Pauli products, $\hat{R}(\theta) = \exp(\alpha \, i \, \theta \, \bigotimes_j \hat \sigma_j)$, by applying each $\hat{\sigma}_j$ in turn. Since they commute, the order of these operators is insignificant.

Some unitaries admit derivatives which cannot be expressed as a (scalar multiple of a) sequence of unitaries, but can as a non-unitary operation. For example, the derivative of the phase gate,
\begin{align}
    \frac{\mathrm{d}}{\mathrm{d}\theta}
    \begin{pmatrix} 1 & 0 \\ 0 & \exp(\iu \theta)
    \end{pmatrix}
    =
    \begin{pmatrix} 0 & 0 \\ 0 & \iu \exp(\iu \theta) 
    \end{pmatrix}
    =
    \iu \exp(\iu \theta) \, |1\rangle\langle 1|,
\end{align}
can be effected by a projection of the target qubit into the $1$ state, and the scaling of $\iu \exp(\iu \theta)$ again deferred to Step \textbf{9} of Algorithm~\ref{alg:state_vec_grad}. Such a projection is likely an existing efficient facility in a simulator, as an embarrassingly parallelisable subroutine used in simulating quantum measurement. 

Analytic unitary matrices specified element-wise can be differentiated either analytically if supported by the simulator and language, else through finite-difference techniques. Ultimately, only a numerical form of the matrix and its derivative, for the current assignment of the ansatz parameters, are needed in Algorithm~\ref{alg:state_vec_grad}. For example, given a gate specified by matrix $\hat{U}_{ij}(\theta) = f_{ij}(\theta)$, where $f$ is a callable function returning a complex scalar, and given the current assignment of $\theta=\phi$, only scalars $\hat{U}_{ij}(\phi)$ and $\frac{\mathrm{d}}{\mathrm{d}\theta}\hat{U}_{ij}(\theta)|_{\theta=\phi} \approx \frac{1}{\delta \theta}(f_{ij}(\phi + \delta \theta) - \hat{U}_{ij}(\phi))$ for each $i,j$ are needed.

Derivatives of \textit{controlled} gates can be effected as above, with an additional step of annihilating state-vector amplitudes of the control qubits, which we notate below as $c$. This is because
\begin{align}
    \frac{\mathrm{d}}{\mathrm{d}\theta} \; C_c[ \hat U(\theta) ]
    &=
    \frac{\mathrm{d}}{\mathrm{d}\theta} \left( 
        ( \hat U(\theta) - \mathbbm{1}) \otimes 
        \ket{1 \dots 1}\!\bra{1\dots 1}_c
        +
        \mathbbm{1} \otimes \mathbbm{1}_c
    \right) 
    \\
    &= \frac{\mathrm{d} \hat{U}(\theta)}{\mathrm{d}\theta}
    \ket{1 \dots 1}\!\bra{1\dots 1}_c.
\end{align}
Hence after applying $\mathrm{d} \hat{U}(\theta)/\mathrm{d}\theta$ to its target qubits, we overwrite amplitudes for which the control qubits are not all $1$, with value $0$.
This is a multi-qubit (but still embarrassingly parallelisable) extension of the projection invoked in the derivative of a phase gate.

An alternative possibility to compute the derivative of a controlled parameterised gate is to decompose it in terms of parameterised single-qubit rotations and CNOTs.
Howver, since the decomposition will generally contain multiple occurences of the gate parameter the product rule must be employed to evaluate the derivative, as
discussed in Section~\ref{app:repeat_param}.


\section{Extensions to state-vector simulation}
\label{app:state_vec_extensions}

    \subsection{Gates with multiple parameters}
    \label{app:multi_param}
    
    Here we outline how to handle the case where a single ansatz gate $\hat{U}_i$ features multiple parameters, $\vec{\phi}$. 
    This is very simple, since the precondition for evaluating the derivative with respect to one parameter of the gate is the same for all its parameters.
    That is, one evaluates all $\nabla \braket{E}$ elements associated with $\vec\phi$ at the stage of Algorithm~\ref{alg:state_vec_grad} when gate $\hat{U}_i$ is visited.
    We outline this subroutine in Algorithm~\ref{alg:multi_param_gate}.
    
    {\centering 
\begin{minipage}{.7\linewidth}
\begin{algorithm}[H]
\DontPrintSemicolon
\SetKwInOut{Input}{Input}
\SetKwInOut{Output}{Output}


\vspace{.5em}
\tcp{loop over each parameter in gate $\hat{U}_i$}
\For{$j \in \{1, \dots, n\}$} 
  {

    $\ket{\mu} \coloneqq \ket{\phi}$
    \tcp*{clone state in $\mathcal{O}(G)$}
    
    $\ket{\mu} \leftarrow (\mathrm{d} \hat{U}_i / \mathrm{d}\phi_j ) \ket{\mu}$ 
    \tcp*{apply non-unitary in $\mathcal{O}(G)$}
    
    $\nabla\braket{E}_{k_j} = 2 \; \Re \; \braket{\lambda | \mu}$
    \tcp*{compute inner product in $\mathcal{O}(G)$}

  } 
\caption{A replacement of lines \textbf{6-9} in Algorithm~\ref{alg:state_vec_grad} to handle a gate $\hat{U}_j$ with multiple parameters, $\phi_1, \dots, \phi_n$, which correspond to gradient elements with indices $k_1, \dots, k_n$}
\label{alg:multi_param_gate}
\end{algorithm}
  \end{minipage}
  \par }

    \subsection{Repeated parameters}
    \label{app:repeat_param}
    
    Here we outline how to relax the condition of parameter uniqueness between gates, so that a parameter $\theta$ can appear in multiple gates. This does not compromise the performance, which remains dominated by a linear scaling in the number of gates present in the circuit. Note though that in principle, if every parameter appeared at least $P$ times each, then our scheme below becomes as inefficient as finite-difference. However, this is an unrealistic scenario; even deliberate efforts to reduce the number of parameters in an ansatz circuit by repeating them between gates, when permitted by the problem, yield fewer than $P$ repetitions~\cite{cai2020resource}. 
    
    Assume $\theta$ is present only in gates $\hat U_i$ \textit{and} $\hat{U}_j$ (where $j>i$). Then by the chain rule,
    \begin{align}
    \frac{\partial\braket{E}}{\partial \theta}
        = \;\;\; & 2 \; \Re \;
    			\bra{\text{in}} 
    			\hat U^\dagger_1
    			\dots 
    			\hat U^\dagger_P
    			\;
    			\hat{H} 
    			\;
    			\hat U_P
    			\dots
    			\hat U_j
    			\dots 
    	    	\frac{\mathrm{d}\hat{U}_i}{\mathrm{d} \theta}
    			\dots 
    			\hat U_1
    			\ket{\text{in}}
    			\\
    	+ \; & 2 \; \Re \;
    		\bra{\text{in}} 
    			\hat U^\dagger_1
    			\dots 
    			\hat U^\dagger_P
    			\;
    			\hat{H} 
    			\;
    			\hat U_P
    			\dots \frac{\mathrm{d}\hat{U}_j}{\mathrm{d} \theta}
    			\dots 
    	    	\hat U_i
    			\dots 
    			\hat U_1
    			\ket{\text{in}}
    \end{align}
    The repetition introduces a new term to $\theta$'s derivative, of the form of that of a unique parameter. If we substitute $U_i(\theta) \to U_i(\phi_1)$ and $U_j(\theta) \to U_j(\phi_2)$, then we can evaluate $\partial\braket{E}/\partial \phi_1$ and $\partial\braket{E}/\partial \phi_2$ independently, and later compute $\partial\braket{E}/\partial \theta = \partial\braket{E}/\partial \phi_1 + \partial\braket{E}/\partial \phi_2$.
    
    We can do this for any number of parameters repeated any number of times; assign all repeated parameters a new unique variable, compute the energy gradient via Algorithm~\ref{alg:state_vec_grad}, then combine elements of the gradient which originally corresponded to the same parameter.
    We outline this process in Algorithm~\ref{alg:repeat_param}.
    
{\centering 
\begin{minipage}{.7\linewidth}
\begin{algorithm}[H]
\DontPrintSemicolon
\SetKwInOut{Input}{Input}
\SetKwInOut{Output}{Output}

\Input{Some representation of a circuit with a single, but possibly repeated, parameter in each gate}
\Output{Energy gradient $\nabla \braket{E}$}

\vspace{.5em}
\tcp{make each parameter unique}
\textbf{let} $ n = P$ 

\textbf{let} $m$ be an empty map

\For{gate in circuit} 
  {
    \textbf{let} gate's parameter be $\theta_i$
  
    \If{$\theta_i$ was already encountered}
    {
    	substitute $\theta_i \to \theta_{n+1}$
    	
    	record $\theta_{n+1} \to \theta_i$ in map $m$
    	
    	increment $n$
    }
    
    record $\theta_i$ as encountered
  } 
 
\vspace{.5em}
\tcp{ensure parameters match gate order}
record ordering of all parameters as $r$

sort all parameters by order of appearance in gates

\vspace{.5em}
\tcp{compute the gradient via new parameters}
compute length-$n$ $\nabla \braket{E'}$ via Algorithm~\ref{alg:state_vec_grad}

reorder $\nabla \braket{E'}$ by ordering $r$

produce length-$P$ $\nabla \braket{E}$ by summing elements of $\nabla \braket{E'}$ according to map $m$
  
\caption{Calculating the gradient under a circuit with repeated parameters}
\label{alg:repeat_param}
\end{algorithm}
  \end{minipage}
  \par }

    \subsection{Non-unitary gates}
    \label{app:non_unitary_ansatz}
    
        Here we describe how to support a non-unitary but invertible ansatz circuit, even one that is not norm-preserving. This introduces only a minor complication in how gates are ``undone" in Algorithm~\ref{alg:state_vec_grad}.
        First, we clarify the role of the conjugate-transpose.  Line~\textbf{11} of Algorithm~\ref{alg:state_vec_grad}, whereby $\ket{\lambda} \leftarrow U_j^\dagger \ket{\lambda} = U_j^\dagger \, U_{j::P}^\dagger \, \hat{H} \, \hat{U}_{1::P} \ket{\text{in}}$, updates state $\ket{\lambda}$ by the \textit{adjoint} of gate $U_j$ only so that in the subsequent inner-product of line~\textbf{9}, this conjugation is undone. That is, line~\textbf{9} computes
        \begin{align}
            \braket{\lambda | \mu} 
            &=
            \bra{\text{in}}  U_{P::1}^\dagger 
            H 
            U_{P::j+1} \; U_j \;
            \ket{\mu}.
        \end{align}
        Hence, the adjiont was invoked because $(U^\dagger)^\dagger = U$, and \textit{not} because $U^\dagger = U^{-1}$. Ergo line~\textbf{9} is unchanged for a non-unitary gate $U_j \to M_j$. In contrast however, \textbf{line}~6 of Algorithm~\ref{alg:state_vec_grad} updates $\ket{\phi} \leftarrow U_j^\dagger \ket{\phi} = U_j^\dagger \hat U_{1::j} \ket{\text{in}}$ with the purpose of \textit{undoing} $U_j$ from the state. Hence to facilitate a non-unitary gate $M_j$, the line should instead read
        \begin{align}
            \ket{\phi} \leftarrow M_j^{-1} \ket{\phi}.
        \end{align}
        For the arguably most common case, where $M_j$ is a single qubit gate, the inverse of its $\mathbb{C}^{2 \times 2}$ matrix can be calculated analytically as
        \begin{align}
            \begin{pmatrix}
            a & b \\ c & d
            \end{pmatrix}^{-1}
            &=
            \frac{1}{ad - bc}
            \begin{pmatrix}
             d & -b \\ -c & a
            \end{pmatrix}.
        \end{align}
        The inverse of a general non-unitary $M_j$ could be found via a generic complex matrix inversion routine, though this may be beyond the facilities of some simulators. 
        
    \subsection{Non-Hermitian operators}
    \label{app:non_herm_op}

    Here we outline a simple extension to Algorithm~\ref{alg:state_vec_grad} to substitute Hamiltonian $\hat H$ with any non-Hermitian operator $\hat A$. This will mean, in general, that the expected value and its gradient are complex~\cite{non_hermitian_expec_valus}. 
    To do so, we must revisit the derivation which assumed Hermiticity in Equation~\ref{eq:deriv_h_hermitian_step}. We have
    \begin{align}
    \frac{\partial\braket{E}}{\partial \theta_i} 
    &=
    \bra{\text{in}} \frac{\partial \hat U^\dagger(\vec\theta)}{\partial \theta_i} \hat A \hat U(\vec\theta) \ket{\text{in}}
    +
    \bra{\text{in}} \hat U^\dagger(\vec\theta)  \hat A \frac{\partial \hat U(\vec\theta)}{\partial \theta_i} \ket{\text{in}}
    \\
    &=
    \bra{\text{in}} \frac{\partial \hat U^\dagger(\vec\theta)}{\partial \theta_i} \hat A \hat U(\vec\theta) \ket{\text{in}}
    +
    \bra{\text{in}}  \frac{\partial \hat U^\dagger(\vec\theta)}{\partial \theta_i}
     \hat{A}^\dagger
    \hat{U}(\vec\theta)  
    \ket{\text{in}}^*,
    \end{align}
    and since $\hat A \ne \hat{A}^\dagger$ in general, we cannot simplify further.
    Instead, we can run Algorithm~\ref{alg:state_vec_grad} (with a minor change) \textit{twice}; once with operator $\hat{A}$ and once with $\hat{A}^\dagger$, and sum their output gradients (complex conjugating the latter) as a final step. The minor change is to replace Line~\textbf{9} of Algorithm~\ref{alg:state_vec_grad} with $\nabla \braket{E}_i = \braket{\lambda | \mu}$, which can now be complex.

\section{Extension to density-matrix simulation}
\label{app:dens_matr}

We do not presently present an algorithm of similar speedup for full-state density matrix simulation, nor do we prove it impossible. Such an extension seems non-trivial, and the density matrix formalism does not permit the same analytic forms leveraged by the state-vector simulation. Instead, we present several intuitive strategies for a density matrix scheme which highlight the difficulty and ultimately do not offer a factor $P$ speedup, but may inspire new directions of optimisation.

The noise-free picture \textit{does} permit an efficient density-matrix algorithm, for a strictly Hermitian operator (e.g. a Hamiltonian as here notated). For $\braket{E(\vec\theta)} = \text{Tr}( \hat{U}(\vec\theta) 
\rhoin{} \hat{U}^\dagger(\vec\theta) \; \hat{H}) $, an element of the energy gradient takes the form
\begin{align}
    \frac{\partial \braket{E}}{\partial \theta_i} 
    & = \text{Tr}\left( \frac{\partial \hat{U}}{\partial \theta_i} \rhoin \hat{U}^\dagger \, \hat{H} + 
    \hat{U} \rhoin 
    \frac{\partial \hat{U}^\dagger}{\partial \theta_i} \, \hat{H}
    \right) \\
    &= \text{Tr}\left( \Lambda_i \hat{H} + \Lambda_i^\dagger \hat{H} \right)
    \tag{assigning $\Lambda_i = \hat{U} \rhoin 
    \frac{\partial \hat{U}^\dagger}{\partial \theta_i}$}
    \\
    &= \text{Tr}\left( 
     \Lambda_i \hat{H} \right) + 
     \text{Tr} \left( 
      \Lambda_i \hat{H}
    \right)^*
    \tag{under $\hat{H}^\dagger=\hat{H}$}
    \\
    &=
    2 \; \Re \; \text{Tr} \left( 
    \hat{U} \rhoin 
    \frac{\partial \hat{U}^\dagger}{\partial \theta_i} \, \hat{H}
    \right).
\end{align}
To evaluate this iteratively, we could follow a similar protocol to the state-vector algorithm, by maintaining numerical representations of $\hat{U} \rhoin$ and $\partial \hat{U}^\dagger/\partial \theta_i \hat{H}$ separately, where we've prior expanded $\hat{H}$ into a dense $2^N \times 2^N$ complex matrix (a one-time overhead, so that we can treat it like a state). The trace is an analogue of the inner-product, and can be evaluated at the same $\mathcal{O}(2^{2N})$ cost as applying a gate (avoiding the otherwise expensive expensive full matrix product) by leveraging that	\begin{align}
			     \text{Tr}(\mu \, \eta) = \sum\limits_i^{2^N} (\mu \, \eta)_{ii}
			    = \sum\limits_j^{2^N} \sum\limits_k^{2^N} \mu_{jk}\,\eta_{kj}.
			\end{align}
However, the introduction of noise operators disables this strategy. Let a channel $\mathcal{D}_i(\rho) = \sum_{j_i} \hat{K}_{j_i} \rho \hat{K}_{j_i}^{\dagger}$ composed of Kraus operators $\{ \hat{K}_{j_i}: j_i = 1, 2, \dots \}$ follow each unitary gate $\hat U_i$ in the ansatz circuit. We assume the noise is independent of the parameters, which otherwise introduces additional terms in the expressions below.
We can first appreciate that ``undoing" an operator from a register, which previously exploited that $\hat{U}^{-1} = \hat{U}^\dagger$, is not so simple for a general Kraus map. Though possible for strictly invertible noise (e.g. Pauli channels), general channels cannot be undone and require caching the state before operation in order to later restore it; this introduces already an $\mathcal{O}(P)$ memory cost. 
Furthermore, the presence of multiple terms in each channel jeopardises an efficient recurrent scheme. 
In this picture, the $i$th element of the gradient is
\begin{align}
\frac{\partial \braket{E}}{\partial \theta_i}
	&=
	2 \; \Re \; \text{Tr} \left( 
		\hat{H} \; 
		\sum\limits_{j_P} \dots \sum \limits_{j_1} \;
		\hat{K}_{j_P} \hat U_P \; \dots \; 
		\hat{K}_{j_i} \frac{\mathrm{d}\hat{U}_i}{\mathrm{d}\theta_i}
		\; \dots \;
		\hat{K}_{j_1} \hat{U}_1
		\; \rhoin \; 
		\hat{U}_1^\dagger
		\hat{K}_{j_1}^\dagger 
		\; \dots \;
		\hat{U}_P^\dagger
		\hat{K}_{j_P}^\dagger 
	\right).
	\label{eq:app_dens_deriv_kraus_form}
\end{align}
Observe there is now no natural partition between two sub-expressions. Instead, there are $\mathcal{O}(M^P)$ terms (where $M$ is the maximum number of Kraus operators per channel), each of which would need to be separately maintained in any iterative evaluation. 
This is easily intuited; in the state-vector picture, unitaries could be applied ``in reverse" by \textit{left}-multiplying them onto the adjoint space. In the density matrix picture however, applying noise in reverse would require an ``outward in" evaluation of the operators in Equation~\ref{eq:app_dens_deriv_kraus_form}. This cannot be done without deferring evaluation until the inner most channel is evaluated, and hence requires propagating $\mathcal{O}(M^P)$ terms.

Another idea is to invoke the Choi-Jamiolkowski isomorphism; we replace the $2^N \times 2^N$ matrix $\rhoin$ with a $2^{2N}\times 1$ vector which we notate as $\ket{\ket{\text{in}}}$. We transform
\begin{align}
    \hat{U} \rhoin \hat{U}^\dagger &\to U^* \otimes U \ket{\ket{\text{in}}},
\end{align}
which admits operation in the same time, since $\hat U$ and $\hat U^*$ can be applied separately to different partitions of the state without evaluation of the tensor product, in $\mathcal{O}(2^{2N})$.
The benefit is that each Kraus map can be replaced with a single superoperator,
\begin{align}
S = \sum\limits_j \hat{K}_j^* \otimes \hat{K}_j,
\hspace{1.5cm}
\sum\limits_j  \hat{K}_j \rho \, \hat{K}_j^{\dagger}
\to 
S \ket{\ket{\rho}}
\end{align}
so that the gradient now has the form
\begin{align}
\frac{\partial \braket{E}}{\partial \theta_i}
&=
    2 \; \Re \; \text{trace}\left[ 
    \left( \mathbbm{1}^{\otimes 2^N} \otimes \hat{H} \right) 
    \; S_P \, (U_P^* \otimes U_P) \dots S_i \frac{\mathrm{d} }{\mathrm{d} \theta_i} (U_i^* \otimes U_i) \dots 
	S_1
	(U_1^* \otimes U_1)
	\;
	\ket{\ket{\text{in}}}
    \right],
\end{align}
where 
\begin{align}
    \text{trace} \,[ \, \ket{\ket{\rho}} \,] = \sum_{n=0}^{2^N-1} \ket{\ket{\rho}}_{(2^N+1)n}.
\end{align}
This new form \textit{does} permit iterative evaluation similar to that in Algorithm~\ref{alg:state_vec_grad}, but the cost of each operator when ``applied left" has become quadratically more costly. 
To illustrate this, consider that
right-applying $S$ onto $\ket{\ket{\rho}}$ is a multiplication of a mostly-identity-matrix (i.e. one of the form $\mathbbm{1}^{\otimes} \otimes M \otimes \mathbbm{1}^{\otimes}$ where $M$ is a fixed size)
onto a $2^{2N} \times 1$ vector, and hence costs $\mathcal{O}(2^{2N})$.
However, an iterative scheme analogous to that in Algorithm~\ref{alg:state_vec_grad} would require we first populate $(\mathbbm{1}^{\otimes 2^N} \otimes \hat{H})$ as a dense $2^{2N} \times 2^{2N}$ matrix (already costing $\mathcal{O}(2^{4N})$), and left-apply operators upon it. Even if this can be done efficiently to make use of $S$ being mostly-identity (and a strategy to do so is not obvious), it costs at least quadratically more per-gate than in a naive evaluation of the gradient.
It appears an iterative scheme using the Choi-Jamiolkowski would cost $\mathcal{O}(P 2^{4N})$, an expectedly much steeper cost than a naive gradient evaluation of $\mathcal{O}(P^2 2^{2N})$, since otherwise the parameters $P$ in an ansatz would exceed that needed for complete description of any state.

\section{Benchmark specifications}

The runtime statistics have been generated with Qiskit 0.20.0 on a 3.1 GHz Dual-Core Intel i7 machine with 16 GB of RAM, running MacOS 10.15.5.
  
\end{document}